\documentclass[aps,prb,preprint,superscriptaddress,showpacs,showkeys]{revtex4}

\usepackage{graphicx}

\usepackage{amssymb}

\begin{document}



\title{Tunneling magnetoresistance of Fe/ZnSe (001) single-  
and double-barrier junctions as a function of interface 
structure}

\author{J. Peralta-Ramos}
\email{Electronic address: peralta@cnea.gov.ar}
\affiliation{Departamento de F\'isica, Centro At\'omico
Constituyentes, Comisi\'on Nacional de Energ\'ia At\'omica,
Buenos Aires, Argentina}
\author{A. M. Llois}
\affiliation{Departamento de F\'isica, Centro At\'omico
Constituyentes, Comisi\'on Nacional de Energ\'ia At\'omica,
Buenos Aires, Argentina}
\affiliation{Departamento de F\'isica, Facultad de 
Ciencias Exactas y Naturales, Universidad de Buenos Aires, 
Buenos Aires, Argentina}

\begin{abstract}
In this contribution, we calculate the spin-dependent ballistic and 
coherent  
transport 
through epitaxial Fe/ZnSe (001) simple and double magnetic 
tunnel junctions with 
two different interface terminations: Zn-terminated and Se-terminated.  
The electronic structure of the junctions is modeled by a second-nearest 
neighbors {\it spd} tight-binding Hamiltonian parametrized to {\it ab initio} 
calculated band structures, while the conductances and the tunneling 
magnetoresistance are calculated within Landauer's formalism. 
The calculations are done at zero bias voltage and as a function of 
energy. We show 
and discuss the influence of the interface structure on the 
spin-dependent transport through simple and double tunnel junctions.
\end{abstract}



\maketitle

\section{Introduction}
\label{intro}

Mixed ferromagnetic/semiconductor nanostructures are gaining 
an increasing interest 
due to their potentialities for spintronic applications, such 
as magnetic random access memories and magnetic field sensors 
(see, for example, Ref. \cite{revs}). It is known that the 
magnetic 
and electronic  
properties of these heterostructures depend on the nature of the 
metal/semiconductor interface, which affects their 
magnetotransport 
properties 
in a significant way \cite{int}. 

For example, M. Eddrief {\it et al} 
\cite{bonding}
have recently 
shown in photoemission experiments that the 
Fe $\Delta_1$ spin-up 
band along the (001) direction, 
which is the band that couples most efficiently to the ZnSe complex bands 
in Fe/ZnSe (001) magnetic tunnel junctions (MTJs) \cite{butler}, 
is strongly modified due to Zn interdiffusion into the Fe electrodes, and that 
these modifications may be the origin of the very low 
tunneling magnetoresistance 
(TMR) measured in these systems. On the theoretical side, 
M. Freyss {\it et al} \cite{freyss}
have shown that the spin polarization $P_S=$($N_{up}-N_{dn}$)/
($N_{up}+N_{dn}$) (being $N_{up/dn}$ 
the density of states of the majority/minority electrons) 
at the Fermi level  
of Fe/ZnSe (001) interfaces is more negative for the Zn-terminated 
interface than for the Se-terminated interface, and 
the transport calculations on Fe/ZnSe (001) single-barrier 
junctions done by H. C. Herper  
{\it et al} \cite{herper} clearly indicate that the TMR 
is higher for Se-terminated interfaces than for Zn-terminated ones. 
These examples 
concerning Fe/ZnSe hybrid nanostructures, together with 
those of Ref. \cite{int} for other hybrid junctions,  
demonstrate that the metal/barrier interfaces play a major role in determining 
the magnetotransport properties of MTJs.

In this contribution, we analyse the transport properties of  
some examples of Fe/ZnSe (001) double-barrier 
tunnel junctions (DMTJs) as compared to 
MTJs, as a function of 
interface termination. To do this, we calculate the 
conductances and the TMR 
of epitaxial Fe/ZnSe (001) tunnel junctions with 
two different interface structures: 
(a) Zn-terminated (both interfaces contain Fe and Zn atoms), and (b) 
Se-terminated (both interfaces contain Fe and Se atoms). 
The results obtained could be  
relevant for the practical use of MTJs and of DMTJs in 
spintronic devices.
\section{Systems under study and calculation methods}
\label{sys}
Our single-barrier magnetic tunnel junctions 
consist of {\it n} layers of zincblende 
ZnSe (001) sandwiched by two BCC Fe (001) semi-infinite electrodes. 
We form DMTJs by inserting {\it m} layers of BCC Fe (001) in between 
the 2{\it n} layers of ZnSe, so that the Fe midlayers are sandwiched by 
{\it n} identical ZnSe layers at each side. In what follows, 
we will call the {\it active region} (AR) to whatever is 
sandwiched by the left and right semi-infinite Fe electrodes. 
Each ZnSe layer and 
Fe midlayer has a width of 0.567 nm  
and 0.287 nm, respectively. The junctions are periodic in the 
{\it x-y} plane, being {\it z} the transport direction. Fig. 1 shows, 
as an example, the interface structure of a Zn-terminated 
MTJ with {\it n}$=$1. The 
Se-terminated junctions are formed by interchanging the Zn and 
the Se atoms. We note that the junctions are epitaxial and that 
interface relaxation is not taken into account. 
 
In the parallel configuration ($P$), the magnetizations 
of all the magnetic regions 
are parallel 
to each other. In the antiparallel configuration ($AP$), 
in MTJs the electrodes' magnetizations are antiparallel 
to each other, while in DMTJs they remain parallel to each other and  
the 
Fe midlayer's magnetization is antiparallel. We note that, since 
the coercive fields of the electrodes and of the midlayer are 
different, these magnetic configurations are experimentally 
achievable \cite{doblese}. 
\begin{figure}[f]
\includegraphics{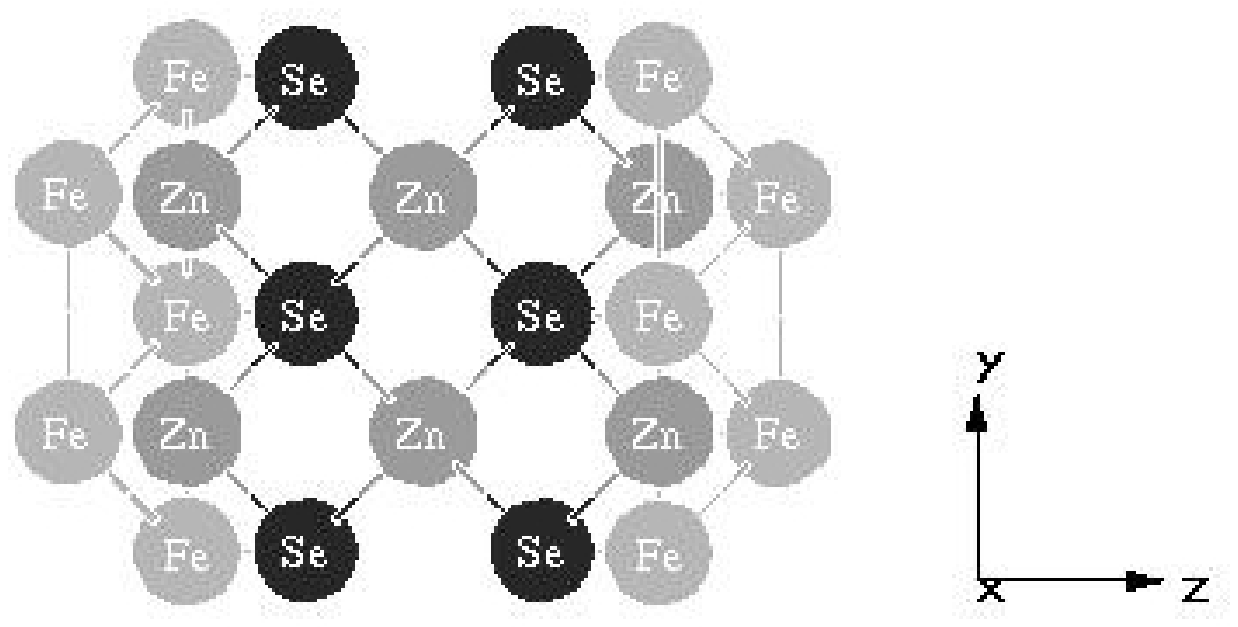}

\caption{Interface structure along the {\it z} direction 
of a Zn-terminated Fe/ZnSe (001) single-barrier junction 
with a ZnSe thickness of 
0.567 nm ({\it n}=1). 
The junction is periodic in the {\it x-y} plane and
the Fe electrodes are semi-infinite.}
\end{figure}

The electronic structure of the junctions is modeled by a second nearest 
neighbors {\it spd} tight-binding Hamiltonian fitted to {\it ab initio}
band structure calculations for bulk Fe and bulk ZnSe \cite{pap}. 
When forming 
the junctions, the ZnSe tight-binding on-site energies are rigidly shifted 
to make the Fe Fermi level fall 1.1 eV 
below its conduction band minimum, as indicated in photoemission experiments 
performed on Fe/ZnSe junctions \cite{edd}. 
Further details can be found in \cite{nosotros}.  

The ballistic conductances $\Gamma$ are calculated using Landauer's 
formalism expressed in terms of Green's functions (see  
\cite{nosotros,datta}). The 
self-energies describing the influence of the electrodes 
on the active region are calculated from the electrodes' 
surface Green's functions (SGFs) in the usual way \cite{datta},  
while the SGFs are obtained using the semi-analytical method described 
in \cite{san}. The tunneling magnetoresistance coefficient is defined 
as TMR=100$\times (\Gamma_P-\Gamma_{AP})/\Gamma_P$, where $\Gamma_P$ 
and $\Gamma_{AP}$ are the conductances in the $P$ and in the 
$AP$ magnetic configurations, respectively. With this 
definition, the TMR ranges from $-\infty$ to 
100 $\%$.
By calculating $\Gamma$ using different numbers
of parallel-to-the-interface wavevectors 
${\bf k_{//}}=k_x {\bf \hat{x}} + k_y {\bf \hat{y}}$ (see Fig. 1), we find that 
5000 is enough to reach convergence. More details on the method used 
to calculate conductances can be found in \cite{nosotros}. 

In this work, we restrict ourselves to zero temperature, 
to infinitesimal bias voltage and 
to the coherent regime (see Ref. \cite{datta}). 
We assume that the electron's ${\bf k_{//}}$ and spin   
are conserved 
during tunneling, since the junctions are epitaxial and 
the Fe midlayer is thin ($<$ 1.8 nm) and 
ordered.
\section{Results and discussion}
\label{res}
Fig. 2 shows the tunneling magnetoresistance of 
Se- and of Zn-terminated MTJs  
with {\it n}$=$2 (1.13 nm), and of DMTJs with 
{\it n}$=$2 and {\it m}$=$6 (1.72 nm), 
as a function of energy (referred to the Fermi 
level $E_F$). 
For single-barrier junctions, 
it is seen that the Se-terminated MTJ has a large TMR,  
near 80 $\%$ on the average, while the Zn-terminated MTJ 
has a much lower one of 40 $\%$, in qualitative 
agreement with the results of \cite{herper}. 
For both terminations, the 
dependence of the TMR on energy is rather smooth. 

For 
double-barrier junctions, it is seen that the
DMTJs' TMR versus energy behaviors are quite different 
for each termination. The Se-terminated double 
junction shows an almost {\it constant}  
TMR enhancement of 
20 $\%$ with respect to the MTJ, 
except for certain energies at which the 
TMR drops abruptly. 
The TMR of 
the Zn-terminated DMTJ is also, in general, larger than the 
corresponding MTJ, but the TMR versus 
energy behavior is not as smooth as in the Se-terminated 
DMTJ.
\begin{figure}[f]
\includegraphics{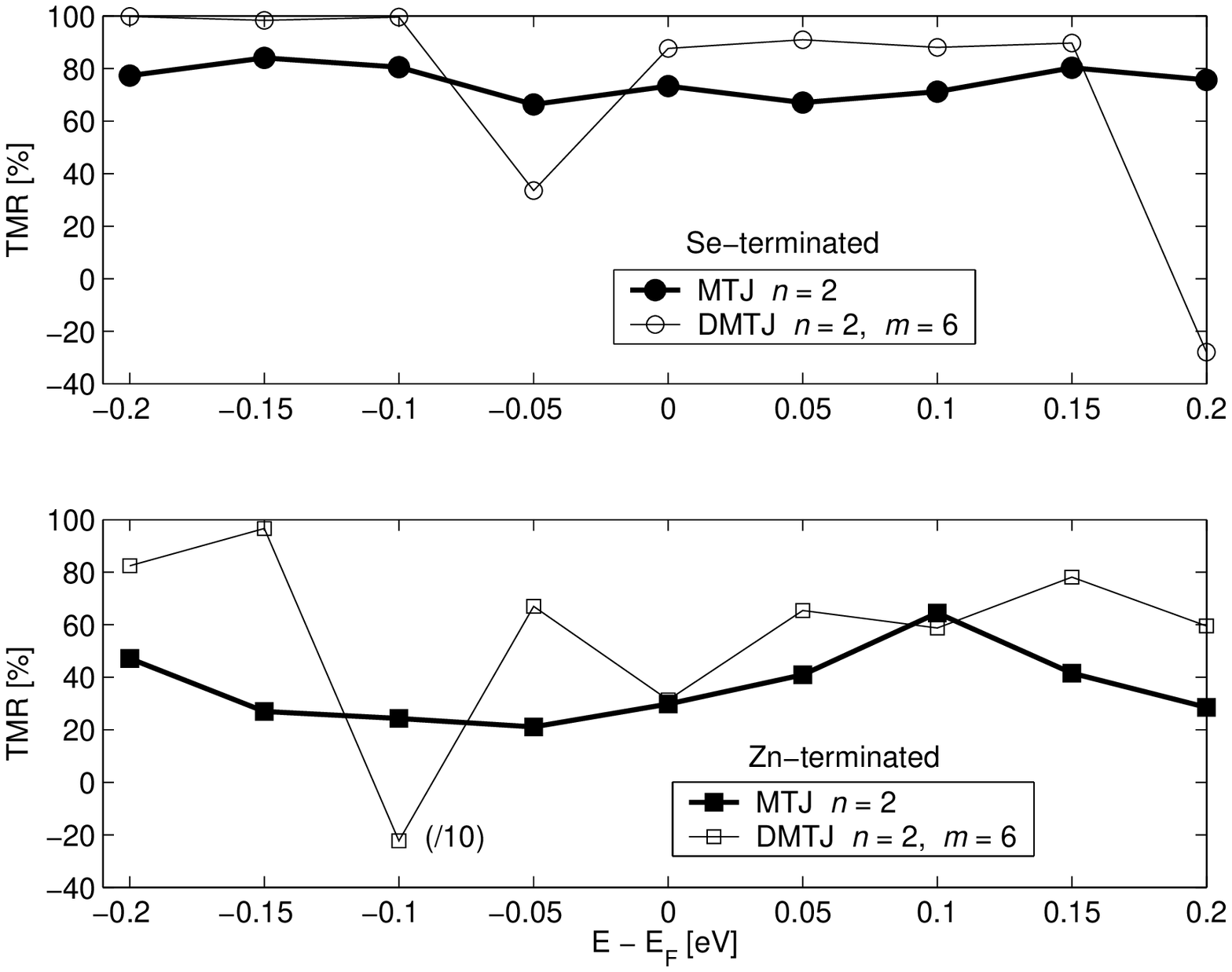}

\caption{Tunneling magnetoresistance as a function of energy 
of MTJs with {\it n}$=$2 (1.13 nm) and of DMTJs  
with {\it n}$=$2 and {\it m}$=$6 (1.72 nm), for 
two different terminations. {\it Upper panel}: Se-terminated. 
{\it Lower 
panel}: Zn-terminated.}
\end{figure}

The TMR versus energy behavior shown in Fig. 2 can be 
understood from Figs. 3 and 4, where we show the 
conductances in the $P$ and $AP$ configurations 
of MTJs (with {\it n}$=$2) and of DMTJs (with {\it n}$=$2 and 
{\it m}$=$6) with Se termination (Fig. 3) and 
with Zn termination (Fig. 4). From the lower panel of 
Fig. 3 (Se-terminated DMTJ) 
it can be observed the $AP$ conductance resonance which 
produces the TMR drop at $E=E_F-$0.05 eV shown in Fig. 2. It  
can also be seen that the conductances 
exhibit an oscillatory behavior with 
energy, indicating the presence of 
resonances. 
Going over to the lower panel of Fig. 4 (Zn-terminated DMTJ) 
it is seen that the TMR drop at $E=E_F-$0.1 eV does not have 
its origin in a resonance in the 
$AP$ conductance but that it is due to a drop in the $P$ 
majority conductance. 
It is also interesting to note that the $P$ majority conductance 
is a smooth function of the energy (except for the already mentioned 
drop), in contrast to the $P$ majority conductance of the Se-terminated 
DMTJ, which shows two peaks for energies 
above $E_F$. These results suggest that, in the Zn-terminated DMTJ that 
we are considering, the 
spin-up quantum well states of the Fe midlayer do not couple to 
the evanescent states in the semiconductor, while in the 
Se-terminated DMTJ they do couple. 
\begin{figure}[f]
\includegraphics{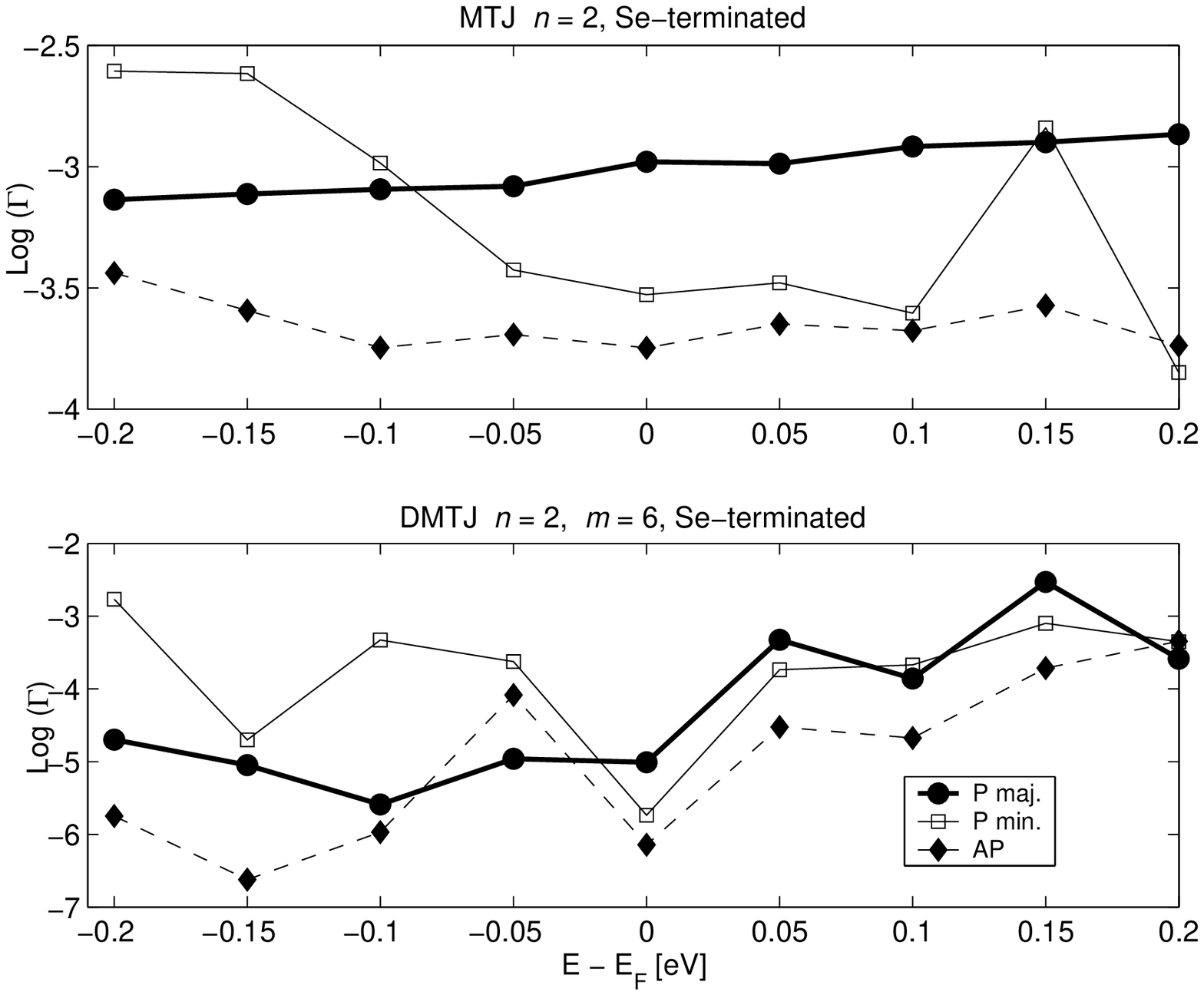}

\caption{Conductances in the $P$ and in the $AP$ configurations 
of a Se-terminated single-barrier junction with {\it n}$=$2  
({\it upper panel}), together with those of a Se-terminated 
double-barrier junction with {\it n}$=$2 and {\it m}$=$6 
({\it lower panel}).}
\end{figure}
\begin{figure}[f]
\includegraphics{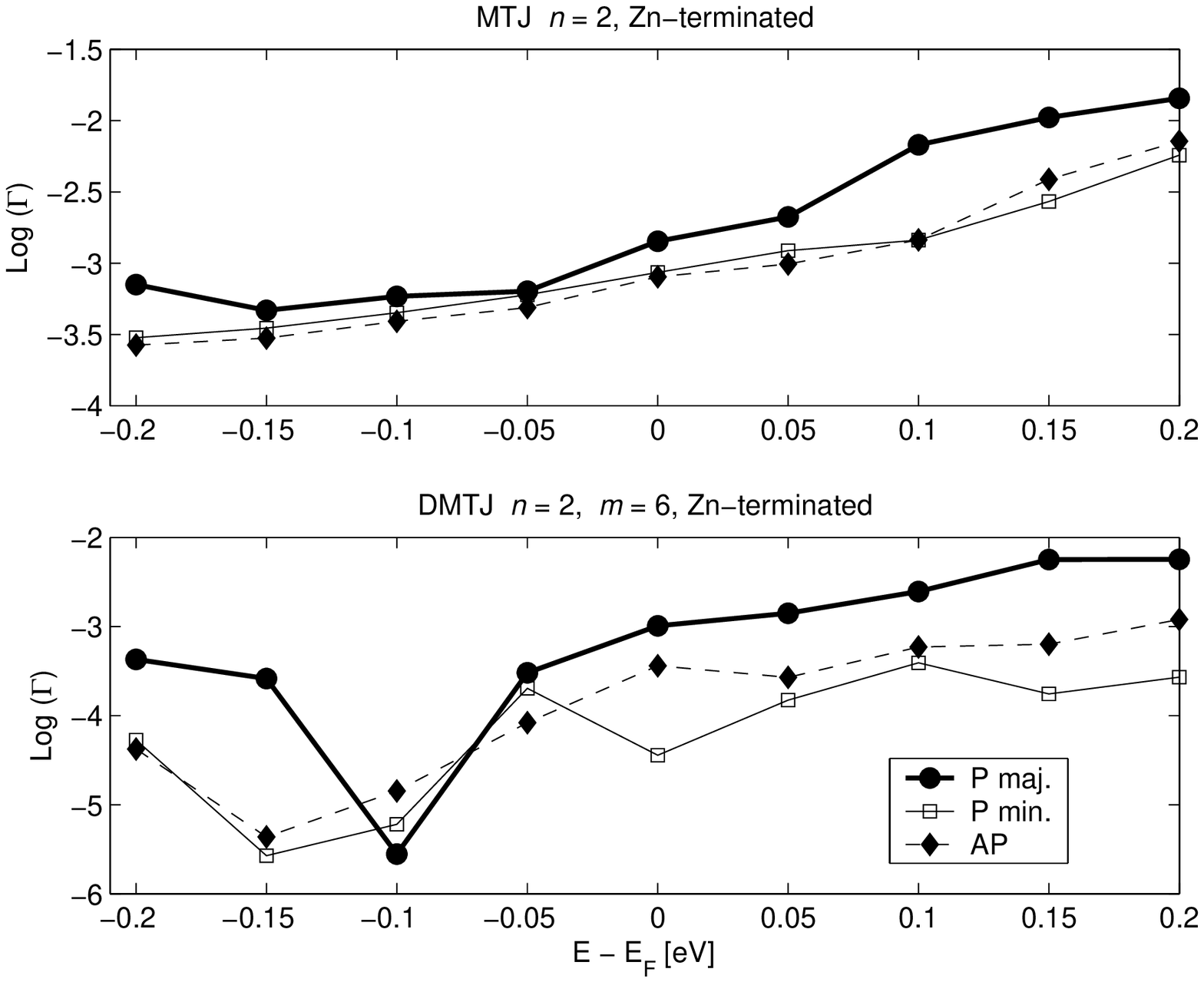}

\caption{Conductances in the $P$ and in the $AP$ configuration 
of a Zn-terminated single-barrier junction with {\it n}$=$2  
({\it upper panel}), together with those of a Zn-terminated 
double-barrier junction with {\it n}$=$2 and {\it m}$=$6 
({\it lower panel}).}
\end{figure}

Another 
interesting feature, that 
can be observed comparing the upper panels of Figs. 3 and 4, is that 
the $P$ minority resonance occurring in the Se-terminated MTJ (Fig. 3), 
does not appear in the Zn-terminated one (Fig. 4). This  
conductance peak is due to the resonant coupling of the well-known 
Fe spin-down 
interface states (which are pinned at $E_F+$0.2 eV \cite{bonding}) 
at 
each interface (see, for example, Refs. 
\cite{int,bonding,butler,freyss}). 
Since it is known that this interface state is  
also present in the Zn-terminated Fe/ZnSe interfaces \cite{freyss}, 
our results indicate that in the Zn-terminated junction the 
interface states at each side of the barrier do not couple to 
each other, while in the Se-terminated case they do. 
We have checked that this 
$P$ minority conductance peak is also present in the 
Se-terminated MTJ with {\it n}$=$4 (2.27 nm), but that it is 
absent in the same MTJ with Zn termination. The $P$ minority 
conductances of the Se- and of the Zn-terminated junctions with 
{\it n}$=$4 are shown in Fig. 5. Note that, as expected, 
the conductance peak in the Se-terminated MTJ 
is not as sharp as that of the one  
with {\it n}$=$2.
We are at present investigating, from first principles, 
the origin of these 
two behaviors, namely the absence of $P$ minority and of $P$ 
majority conductance peaks in the Zn-terminated MTJs and DMTJs, 
respectively.
\begin{figure}[f]
\includegraphics{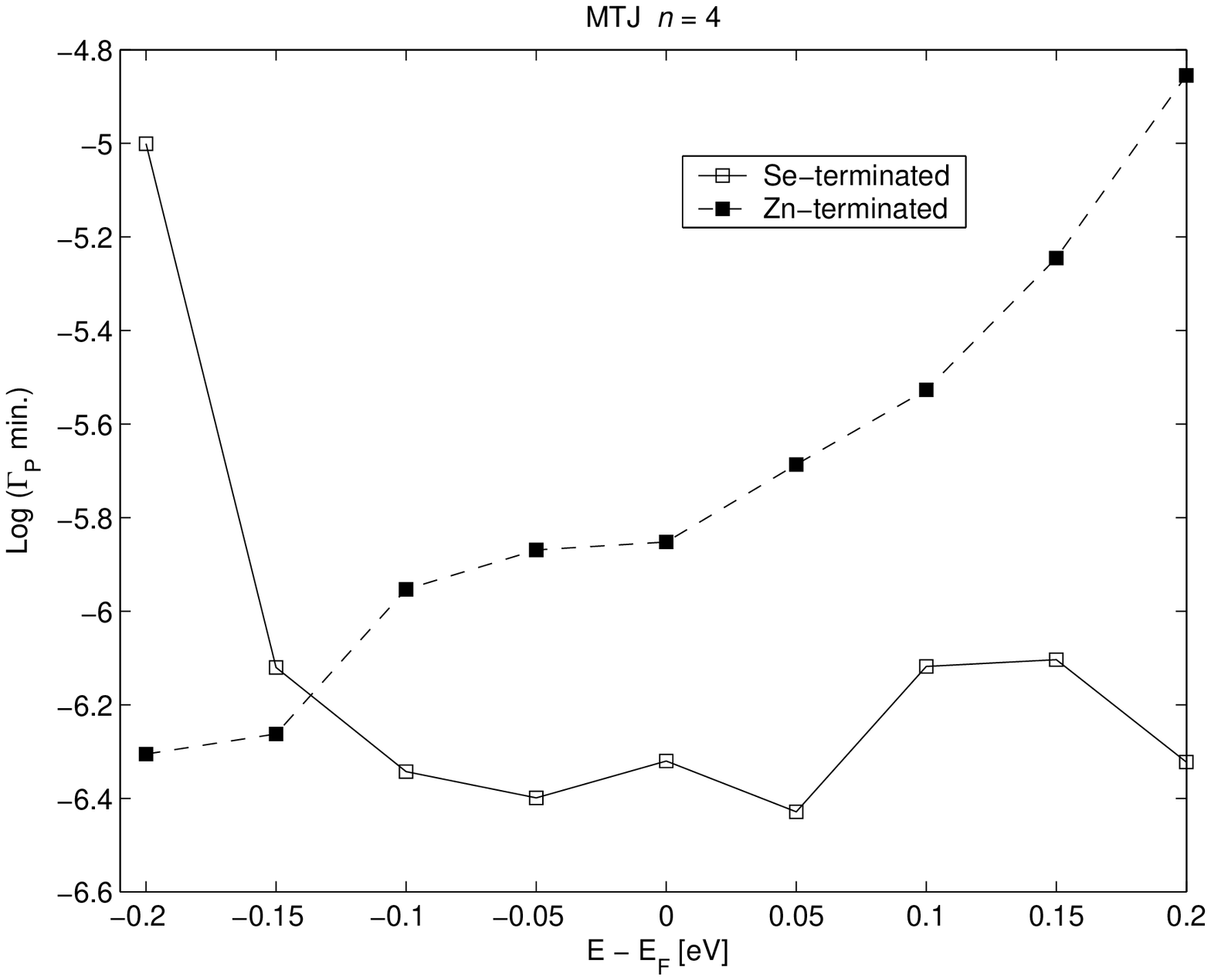}

\caption{Conductance of minority electrons in the parallel configuration 
of Se- and of Zn-terminated 
single-barrier junctions 
with {\it n}$=$4 (2.27 nm).}
\end{figure}

Continuing with the analysis of DMTJs, we show in Table 1 
the TMR values, evaluated at $E_F$ and at $E_F\pm$0.05 eV,  
of DMTJs with {\it n}$=$2 
(1.13 nm) and 
different values of {\it m}, and of the MTJs with {\it n}$=$2 (and 
{\it m}$=$0), for the 
two different interface terminations considered. It is seen that, in 
general, the TMR values of the Se-terminated DMTJs are positive 
and rather large (the maximum is 98.9 $\%$), 
while the ones corresponding to the Zn-terminated 
DMTJs are either positive (but not so large) or 
negative and very large (reaching $-$639 $\%$). These features may result in 
a more pronounced bias voltage dependence of the TMR in 
Zn-terminated DMTJs than in Se-terminated ones, although 
further studies are desirable.
\begin{table}[tbh]
\begin{center}
\begin{tabular}{|c|c|c|}
\hline
{\it m} & Zn-terminated & Se-terminated\\
\hline
0 ({\it n}=2) & 21.1/29.8/40.9 (30.6) & 66.3/73.3/67 (68.9)\\
\hline
2 &$-$639/$-$88.8/27.5 ($-$233.4) & $-$66.1/67.3/87.8 (29.7)\\
\hline
4 & 69.6/$-$98.9/36.2 (2.3) & 92.9/95.9/98.9 (95.9)\\
\hline
6 & 67/31.4/65.4 (54.6) & 33.5/87.7/90.9 (70.7)\\
\hline
\end{tabular}
\end{center}
\vspace{0.2cm}
\caption{TMR values evaluated at $E_F$ and at 
$E_F\pm$0.05 eV of DMTJs with {\it n}$=$2 
(1.13 nm) and different values of {\it m}, 
together with those of the MTJ with {\it n}$=$2 (and 
{\it m}$=$0), for two different interface structures. TMR values 
are given in $\%$ and refer to $E_F-$0.05 eV/$E_F$/$E_F+$0.05 eV. 
The number 
in brackets is the TMR value averaged over the three energies considered.}
\end{table}

\section{Summary}
\label{sum}
Using a realistic model for the electronic structure 
and accurate conductance 
calculations in the coherent, zero bias and ballistic 
regime, we showed that the spin-dependent transport 
properties of Fe/ZnSe (001) single- and 
double-barrier tunnel junctions are very sensitive 
to the Fe/ZnSe interface structure. In particular, we found 
that in the Se-terminated single-barrier junctions considered, 
the  
conductance of minority electrons has a peak that is absent 
in the Zn-terminated junctions, indicating that 
in the latter case the Fe spin-down interface states at each side 
of the ZnSe barrier  
do not couple to each other.
We also found 
that the tunneling magnetoresistance 
of double-barrier junctions reaches higher values  
in the Se-terminated double junctions than in the Zn-terminated 
ones, in which case the TMR can reach very large and negative 
values.

This work was partially funded by UBACyT-X115, Fundaci\'on 
Antorchas, PICT 03-10698 and PIP-Conicet 6016. A. M. Llois 
belongs to CONICET (Argentina).


\begin{thebibliography}{00}



\bibitem{revs} E. Y. Tsymbal, O. N. Mryasov, and P. R. LeClair 
J. Phys.: Condens. Matter {\bfseries 15}, 109 (2003)
\bibitem{int} C. Tiusan {\it et al}, 
J. Phys.: Condens. Matter {\bfseries 18}, 941 (2006); 
C. Tiusan {\it et al}, Phys. Rev. Lett. {\bfseries 93}, 
106602 (2004); M. E. Eames and J. C. Inkson, Appl. Phys. Lett. 
{\bfseries 88}, 252511 (2006);  
C. Heiliger {\it et al}, Phys. Rev. 
B {\bfseries 73}, 214441 (2006); K. D. Belashchenko, J. Velev, and 
E. Y. Tsymbal, cond-mat 0505348 v1, 13 May 2005; 
N. Papanikolaou {\it et al}, 
Phys. Rev. B {\bfseries 62}, 11118 (2000)
\bibitem{bonding} M. Eddrief {\it et al}, 
Phys. Rev. B {\bfseries 73}, 
115315 (2006)
\bibitem{butler} Ph. Mavropoulos, N. Papanikolaou, and P. H. 
Dederichs, Phys. Rev. Lett. {\bfseries 85}, 1088 (2000); 
J. M. MacLaren {\it et al}, 
Phys. Rev. B {\bfseries 59}, 5470 (1999)
\bibitem{freyss} M. Freyss {\it et al}, 
Phys. Rev. B {\bfseries 66}, 014445 (2002)
\bibitem{herper} H. C. Herper {\it et al}, 
Phys. Rev. B {\bfseries 64}, 
184442 (2001)
\bibitem{doblese} T. Nozaki {\it et al}, 
Appl. Phys. Lett. {\bfseries 86}, 082501 (2005); 
Z. M. Zeng {\it et al}, 
J. Magn. Magn. Mater. {\bfseries 303}, 219 (2006)
\bibitem{pap} D. A. Papaconstantopoulos, {\it Handbook of 
the band structure of 
elemental solids} (Plenum Press, New York, 1986); 
R. Viswanatha, S. Sapra, B. Satpati, P.V Satyam, 
B.N Dev, 
and D.D Sarma, cond-mat 0505451 v1, 18 May 2005
\bibitem{edd} M. Eddrief {\it et al}, 
Appl. Phys. Lett. {\bfseries 81},
4553 (2002)
\bibitem{nosotros} J. Peralta-Ramos and A. M. Llois, Phys. Rev. B 
{\bfseries 73}, 214422 (2006) 
\bibitem{datta} S. Datta, {\it Electronic transport in mesoscopic 
systems} (Cambridge University Press, Cambridge, 1999)
\bibitem{san} S. Sanvito, C. J. Lambert, J. H. Jefferson, and
A. M. Bratkovsky, Phys. Rev. B {\bfseries 59}, 11936 (1999)

\end{thebibliography}
\end{document}